# CD16: a MATLAB-based application for estimating radiation doses from x-ray guided cardiac catheterizations in children.

Richard W Harbron[1], Mark S Pearce[1], Claire-Louise Chapple[2]

[1]Newcastle University, Institute of Health and Society, Sir James Spence Institute, Royal Victoria Infirmary, Queen Victoria Road, Newcastle-upon-Tyne, NE1 4LP

[2]Regional Medical Physics Department, Freeman Hospital, Newcastle-upon-Tyne hospitals NHS trust, Newcastle upon Tyne, UK

## Abstract:
CD16 is a MATLAB-based dosimetry system designed for rapid estimation of organ doses for x-ray guided cardiac catheterization procedures. CD16 is principally designed for epidemiological studies in which dose estimates are required for thousands of cohort members, often based on limited exposure information. The minimum data required for dose estimates are (1) total examination kerma area product ($P_{KA}$), (2) patient size (age or mass), (3) x-ray spectrum represented by half value layer (HVL, in mm Al) and (4) examination type. These data can be inputted using an Excel spreadsheet. Organ dose estimates are outputted into the same spreadsheet. Doses are estimated using lookup tables of *k*-factors relating $P_{KA}$ to organ doses for a wide range of beam angles, six patient sizes and two field sizes. These *K*-factors were calculated using the Monte Carlo code PCXMC V2.0. CD16 initially calculates organ doses for a beam energy represented by a HVL of 4.8 mm Al, then adjusts to other beam energies if specified. Beam angles can be specified if known, otherwise CD16 assigns beam angles based on examination type. CD16 is the final version of this dosimetry system to utilise PCXMC. Future versions will use the general purpose Monte Carlo code MCNP and hybrid phantoms.

## Introduction:

CD16 (short for 'cardiodose' 2016) is a MATLAB-based dosimetry system that allows rapid estimation of mean absorbed dose to nine organs (in mGy), along with effective dose (in mSv) and average absorbed dose to the whole body (mGy) for a range of cardiac catheterization procedure types. CD16 was principally designed for epidemiological analysis, where organ dose estimation may be required for several thousand patients, often based on limited details recorded at the time of examination. Although it would be desirable to be able to reconstruct past exposures based on detailed structured dose reports, for most examinations performed <2010 these data simply aren't available. CD16 requires only limited information for dose estimates:

1. Kerma area product ($P_{KA}$ in cGy*cm$^2$, also referred to as dose area product, DAP)
2. Patient size (mass in kg or age in years)
3. Beam energy (half value layer, HVL, in mm Al).
4. Examination type

Beam angles can be specified if known. If not, they are estimated based on examination type. If this is not known either, the examination is assigned as an 'unspecified cardiac catheterization' using a combination of frontal and lateral beam angles in 0.6/0.4 proportions. If $P_{KA}$ is recorded in two planes (i.e. frontal and lateral), these two figures can be used to give more accurate dose estimates. A table reduction factor to account for beam attenuation by the table and mattress can be applied, if appropriate. Data can be inputted and the results exported in the form of an Excel spreadsheet. CD16 was originally designed for use in paediatric dose estimations. It can, however, be used for all patient ages, although caution is advised for obese patients (see Limitations section).

CD16 is operated as a MATLAB function. As with other dose estimation tools (e.g. [1]) it utilises data in the form of lookup tables derived from Monte Carlo simulations, but does not perform any 'live' Monte Carlo simulations itself. This has the advantage of computational efficiency when estimating doses for a large cohort of patients. The downside of the lookup table approach is a lack of flexibility. To address this, the lookup tables used in CD16 are very large, incorporating an extensive range of beam angles, six patient sizes and two field sizes, with the ability to interpolate between beam angles and patient sizes.

MATLAB is a commercial 4th generation programming language. Most universities will have a licence, especially if there is an engineering or physics department. There are a number of free programs that use the same syntax as MATLAB, including *GNU Octave* and *Scilab*. We have not attempted to use CD16 with either of these, but would be interested to hear anyone's experience of trying.

Lookup table construction:

As a general principle, organ doses ($D$), can be estimated using the following relationship:

$$D = dose\ indicator * k$$

Where $k$ is a beam angle, field size, patient size and beam energy specific conversion factor. The value of $k$ can be determined using either physical measurements in anthropomorphic phantoms, or Monte Carlo (MC) computer simulations. The latter was used in this case, namely PCXMC V2.0 [2]. Dose indicators include kerma area product ($P_{KA}$, also called dose area product) in radiography and fluoroscopy, and dose length product (DLP) and CT dose index (CTDI) in computed tomography. In CD16, $P_{KA}$ is used as the input parameter. The default unit for $P_{KA}$ is Gy·cm$^2$ (equivalent to 100 µGy·cm$^2$).

PCXMC was developed by STUK (Helsinki, Finland) [2], designed specifically for diagnostic x-ray dose estimation. PCXMC simulates photon interactions, but not electrons, thus assumes x-ray energy is deposited locally at the site of interaction. The program uses modified versions of Cristy/Eckerman phantoms in 6 sizes (0, 1, 5, 10, 15 and 30 years) (Table 1). Limitations of these phantoms are described later.

Table 1: Phantom characteristics in PCXMC 2.0. Data from Tapiovaara and Siiskonen [2].

| Phantom size | Mass (kg) | Total height (cm) | Trunk height (cm) | Trunk thickness (cm) | Lung density (g cm$^{-3}$) | Soft tissue density (g cm$^{-3}$) | Bone density (g cm$^{-3}$) |
|---|---|---|---|---|---|---|---|
| New born | 3.0 | 50.9 | 21.6 | 9.8 | 0.30 | 1.04 | 1.22 |
| 1 Year | 9.2 | 74.4 | 30.7 | 13.0 | 0.30 | 1.04 | 1.40 |
| 5 Years | 19.0 | 109.1 | 40.8 | 15.0 | 0.30 | 1.04 | 1.40 |
| 10 Years | 32.4 | 139.8 | 50.8 | 16.8 | 0.30 | 1.04 | 1.40 |
| 15 Years | 56.3 | 168.1 | 63.1 | 19.6 | 0.30 | 1.04 | 1.40 |
| Adult | 73.2 | 178.6 | 70.0 | 20.0 | 0.30 | 1.04 | 1.40 |

PCXMC can be used to obtain organ dose estimates for a given input parameter ($P_{KA}$, incident air kerma, mAs, etc.) for any beam angle. The two beam angle directions are 'Projection angle' and 'Cranio-caudal angle (Figures 1 and 2). In PCXMC, a negative cranio-caudal angle means the beam is angled towards the head. A positive cranio-caudal beam angle is angled towards the feet. The 'Projection angles' used in PCXMC are not necessarily the same as those quoted by equipment used in clinical practice. A projection angle of 90° represents the postero-anterior (PA) projection (beam entering the posterior aspect of the patient and exiting the anterior), while 180° represents a left lateral projection (beam entering the right side of the patient and exiting the left). Note that many cardiologists incorrectly refer to the PA projection as 'AP'. True AP (anteroposterior) projections are almost never used in clinical practice.

Figure 1: Cranio-caudal angulation demonstrating PCXMC signed angle terminology. The patient is shown supine with arms raised above the head.

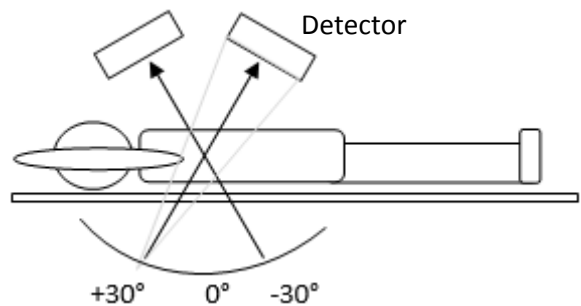

Figure 2: 'Projection angle'. Figure represents looking from the head of a supine patient, towards the feet. The arrows represent the direction of the beam. Note, projection names indicate where the beam *exits* the patient. Thus, 'left lateral' means the beam exits the left side of the patient.

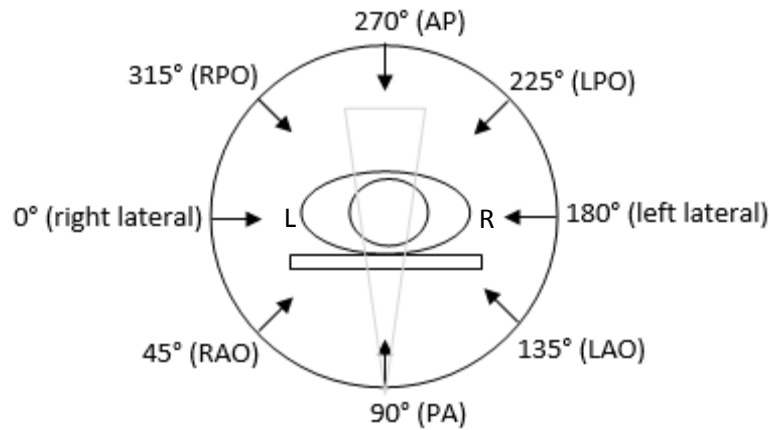

PA = postero-anterior, AP = antero-posterior, LPO = left posterior oblique, RPO = right posterior oblique, RAO = right anterior oblique, LAO = left anterior oblique

Organ doses, per unit $P_{KA}$ were calculated using PCXMC for 72 'projection' beam angles, representing a full 360° rotation around the patient in 5° intervals. The arms were removed for all exposures (in reality, they are raised above the head to allow lateral projections). Doses were calculated at each of the 72 projection angles for 23 different angles in the cranio-caudal direction, ranging from 55° cranial to 55° caudal. This produced a total of 1656 beam angle specific *k*-factors (72 * 23) for each organ, effective and mean whole body dose. Tube potential was initially set at 70 kV, while aluminium and copper filtration levels were set at 2.5 and 0.2 mm, respectively. This resulted in a HVL of ~4.8 mm Al. A beam energy correction factor (described later) is applied to allow dose estimation for other beam spectra. Simulations were run using 500,000 photons. This kept MC simulation uncertainties below 1% for most organs. Uncertainties were greatest for the thyroid, which lies outside the primary radiation field.

This process was repeated for each of 6 patient sizes (Table 1). Two field sizes were used (Figures 3 and 4): (1) a 'large' field, designed to include the major pulmonary vessels and aortic arch, and (2) a 'close' field, in which the beam is collimated to the heart and a small surrounding region. This resulted in a total of 19,872 *k*-factors for each organ type (1656 angles, 6 phantom sizes and 2 field sizes). Only *k*-factors for 9 selected organs are used in CD16. The remainder were organs of little relevance to cardiac exposures, either because they are well outside the exposed region (e.g. brain, bladder, gonads), or due to the lack of observed effects of low dose radiation exposure (e.g. muscle).

Organ dose estimates are required for a range of different beam spectra, i.e. not just at 4.8 mm Al. A beam energy correction factor, for each combination of organ, beam angle, field size and patient size, was calculated by fitting a $4^{th}$ degree polynomial using MATLAB to dose estimates calculated at 9 different values of HVL, relative to 4.8 mm Al. These 9 different HVLs were produced by variation the tube potential from 50 to 100 kV and the added copper filtration from 0 to 0.9 mm. The simulations used to calculate the beam energy correction factor were run at 100,000 photons. Overall, 178,848 simulations were performed; 1656 beam angles, 9 beam energies, 2 field sizes and 6 patient sizes.

### CD16 MATLAB function:

A function was written in MATLAB to pick the right *k*-factor, given the patient size, beam energy and examination type. These input parameters can be inputted in the form of an Excel spreadsheet (described in Operation of CD16).

Specified angles in between the simulated 5° intervals are estimated using linear interpolation. CD16 works by estimating the organ dose for each of the 6 patient sizes, corrects each for beam energy, then estimates doses at the desired patient size by linear interpolation between patient sizes. A power-law interpolation method also was investigated, but was abandoned due to the prediction of negative doses for large patient sizes, or a dose of infinity for new-born babies (i.e. an age of 0 years). The linear interpolation is truncated at 80 kg and 18 years to prevent very low or even negative dose estimates for large patient sizes or ages beyond the end of childhood. CD16 only really works for 'normal' sized patients. Dose estimates for obese patients are likely to be especially unreliable. PCXMC can simulate obese patients, but does so in such an unrealistic way (see Limitations section) that this feature was not utilised.

If beam angles are known, these can be specified used for dose estimation. Otherwise CD16 assigns beam angles depending on procedure type. The beam angles used for each procedure type are encoded in a $n \times 4$ matrix, written in the MATLAB function script as a single row. Matrix columns are separated with a comma and rows are separated with a semicolon.

Each row represents a beam angle, in degrees, in PCXMC language. The first column is the 'projection' beam angle and the second column is the cranio-caudal angle. The third column is the proportion of the total $P_{KA}$ to be applied for the respective angle. The forth column is the table reduction factor. Set this to "t" to apply the standard table reduction (i.e. used for any beam angle that will pass through the table before reaching the patient), and zero otherwise. The value of t can be adjusted to whatever is desired. The default value is 0.2, which means doses are reduced by a factor of 20%.

The default beam angles shown below for each procedure type were determined from three sources: (1) structured dose reports and stored images for a sample of 316 procedures, (2) logbook records of beam angles, and (3) biplane $P_{KA}$ figures.

The structured dose reports provided detailed information on beam angles, tube potential and added copper filtration for digital acquisitions only. No corresponding information was given for fluoroscopic imaging. Images stored on the PACS network were also primarily digital acquisitions, although a number of 'fluoro grabs' (stored frames or sequences of fluoroscopic imaging) were also recorded. Some modern equipment does record beam angles and exposure factors for all exposures, including fluoroscopy. These data could be obtained to improve dose estimates in the future.

Biplane $P_{KA}$ figures account for both fluoroscopic and acquisition exposures, but provide no information concerning tube potential, filtration, or the actual beam angles in each plane. They are useful for examinations involving simple beam angle combinations (i.e. PA and lateral only), such as pulmonary and aortic valvuloplasty, but less useful for examinations with complex, multiple beam angles, such as coronary angiography.

Uses are free to change the default beam angles by editing the CD16 function script.

*Procedure type 0: Unspecified cardiac catheterization:*

$$\begin{bmatrix} 90 & 0 & 0.6 & t \\ 180 & 0 & 0.4 & 0 \end{bmatrix}$$

This means there are two beam angles: One with a rotational angle of 90° (straight PA), the other 180° (straight left lateral). Neither have any cranio-caudal angulation. The third column shows that the proportions of the PA and lateral angles are 0.6 and 0.4, respectively.

Here's how this looks in the MATLAB function script:

$$[90,0,0.6,t; 180,0,0.4,0];$$

For example, for a total examination $P_{KA}$ of 2.0 Gy·cm², 1.2 Gy·cm² would be applied in a beam angle of 90° (straight PA) and 0.8 Gy·cm² would be applied in a beam angle of 180° (straight left lateral).

*Type 1: ASD occlusion*

$$\begin{bmatrix} 90 & 0 & 0.93 & t \\ 180 & 0 & 0.07 & 0 \end{bmatrix}$$

*Type 2: PDA occlusion*

$$\begin{bmatrix} 90 & 0 & 0.3 & t \\ 180 & 0 & 0.6 & 0 \\ 60 & 0 & 0.1 & t \end{bmatrix}$$

*Type 3: Pulmonary valvuloplasty*

$$\begin{bmatrix} 90 & 0 & 0.60 & t \\ 180 & 0 & 0.40 & 0 \end{bmatrix}$$

*Type 4: Aortic valvuloplasty*

$$\begin{bmatrix} 90 & 0 & 0.65 & t \\ 180 & 0 & 0.35 & 0 \end{bmatrix}$$

### Type 5: Pulmonary artery angioplasty

$$\begin{bmatrix} 90 & 0 & 0.25 & t \\ 180 & 0 & 0.4 & 0 \\ 115 & -25 & 0.25 & t \\ 65 & -25 & 0.1 & t \end{bmatrix}$$

### Type 6: Coarctation repair

$$\begin{bmatrix} 90 & 0 & 0.55 & t \\ 180 & 0 & 0.45 & 0 \end{bmatrix}$$

### Type 7: Electrophysiology study (EPS)

$$\begin{bmatrix} 60 & 0 & 0.5 & t \\ 40 & 0 & 0.5 & 0 \end{bmatrix}$$

### Type 8: Endomyocardial biopsy

$$\begin{bmatrix} 90 & 0 & 0.89 & t \\ 180 & 0 & 0.11 & 0 \end{bmatrix}$$

### Type 9: Coronary angiography:

$$\begin{bmatrix} 70 & -40 & 0.2 & t \\ 70 & 40 & 0.2 & t \\ 140 & 30 & 0.2 & t \\ 50 & 0 & 0.2 & t \\ 130 & 0 & 0.2 & t \end{bmatrix}$$

### Type 10: Pressures and pulmonary vascular resistance studies

$$\begin{bmatrix} 90 & 0 & 0.75 & t \\ 180 & 0 & 0.25 & 0 \end{bmatrix}$$

### Type 11: Trans-catheter pulmonary valve replacement

$$\begin{bmatrix} 90 & 0 & 0.65 & t \\ 180 & 0 & 0.35 & 0 \end{bmatrix}$$

### Type 13: Atrial septostomy

$$\begin{bmatrix} 90 & 0 & 0.75 & t \\ 180 & 0 & 0.25 & 0 \end{bmatrix}$$

## Operation of CD16:

To use CD16, six files are needed and should be stored in the MATLAB home directory:

- The CD16 MATLAB code itself
- Reference file for small field size ("refSize1New.mat")
- Reference file for large field size ("refSize2New.mat")
- Energy corrections for small field size ("HVLpolyFS1.mat")
- Energy corrections for large field size ("HVLpolyFS1.mat")
- Input file containing examination data

## Creating the input file:

It is easiest to input the examinations parameters into an Excel file, especially for large cohorts. Each row of the file is an examination, or part of an examination (e.g. a different beam angle). The columns represent the various input parameters.

Columns 1-11 of input file:

| ID | DAP | Proj beam angle | CC beam angle | Mass | HVL | Exam | Field size | PA DAP | Lat DAP | Table |
|----|-----|-----------------|---------------|------|-----|------|------------|--------|---------|-------|

1. ID is just some form of identifier for the row. It could be used for a patient ID or just left blank.
2. DAP is dose area product, or kerma area product. The units default to cGy*cm$^2$ as this is what most machines report. Must be specified.
3. Proj beam angle is the projection angle shown in Figure 2.
4. CC beam angle is the cranio-caudal beam angle shown in Figure 1.
5. Mass is patient mass in kg. Either mass or age must be specified.
6. Age is patient age in years.
7. HVL represents the beam energy in half value layer (mm Al). If left blank, a HVL of 4.8 mm Al is used.

8. Exam is the examination type (1-13, see list above)

9. Field size can be either 1 or 2 for small or large, respectively.

10-11. PA DAP and Lat DAP can be entered if biplane DAP figures are available. If one plane is zero, enter this as such, otherwise CD16 will ignore the biplane information.

12. Table tells CD16 whether or not to apply the table attenuation factor – 0 for no, 1 for yes.

Columns 13-25:

| E | ABM | Breasts | Heart | Lungs | Lymph | Oesoph | Thyroid | Stomach | Mean D | m/f | dataset |
|---|-----|---------|-------|-------|-------|--------|---------|---------|--------|-----|---------|

13-23. Leave these columns blank. They will be filled in with output data from CD16.

24. 'm/f' is not used in CD16. This column was used in risk estimates, which are sex-specific

25. Allows results to be sorted according to hospital, if multi-centre datasets are used.

The bare minimum information for CD16 to work is $P_{KA}$ (DAP, column 2), and a measure of patient size (Mass in column 5 or Age in column 27). If the other fields are left blank, CD16 will use default values. The more information provided, the better the dose estimates will be, however.

## Loading the input file:

The input file should be loaded into MATLAB as a single matrix. There are a number of ways of doing this:

1. Right click on the input file in the MATLAB file browser, then select 'Import Data' from the dropdown menu. This will bring up a new window allowing you to select the cells you want to import. It's best to highlight the entire rows by left clicking and dragging down the left hand column all the way to the end of the input file. Make sure 'matrix' is selected in the toolbar at the top of the window, rather than 'column vectors'. Also make sure the full width of the input file is imported, e.g. columns A to Y, otherwise MATLAB will report an error.

2. Click on the 'Import data' icon on the MATLAB toolbar and find the file, then proceed as above.

3. You can use the 'xlsread' command in MATLAB. For example.
   inputFile=xlsread('inputFile.xlsx', 'Sheet1', 'A2:Y100'); This will load in a matrix called 'inputFile' from the Excel sheet with the same name. This works better when the input file is large, as the Import data function in MATLAB can be very slow in such cases.

## Running CD16:

Once the input file has been created, to calculate doses, just type:

output=CD16(inputFile);

Where "inputFile" is whatever the input file matrix is called.

The function is automatically set up to save the output back into the original input file in columns L to V. Make sure the file is not open in Excel while CD16 is running, otherwise a write error will result. To disable automatic writing of results in to the Excel file, place a "%" symbol in front of the last line of code. This will cause the text to turn green. MATLAB will also return an output matrix as 'ans' in the workspace (any variable name can be used for this).

Upon opening the Excel file containing the input data, columns M to W should be filled with dose estimates in mGy or mSv. If any cells are blank, it is likely MATLAB returned a value NaN (not a number).

For the current UK cardiology cohort, which includes around 20,000 examinations, CD16 takes around 80 minutes to estimate organ doses.

Changes since previous publications:

CD16 is the latest version of a series of similar dosimetry systems. It will be the final version to use PCXMC with Cristy/Eckerman phantoms to derive $k$-factors used in dose estimation. Later versions will use a different Monte Carlo code (MCNP) and more anatomically realistic hybrid phantoms [3].

Earlier versions of CD16 have been used in three previously published studies [4-6]. In the first paper [4], a single field size was used, shown by the red box in Figure 4. For the other two papers [5, 6], two field sizes were used, shown by the red boxes in Figures 3 and 4. Since these papers, a number of modifications were made:

- For the large field size, the z-axis beam centring point was adjusted superiorly (i.e. towards the head), to correspond to the approximate location of the pulmonary valve, rather than the centre of the whole heart. The y-axis centring point was also translated in the posterior direction (i.e. towards the spine). For laterally orientated projections, this latter change results in the complete exclusion of the anterior chest wall from the primary x-ray field.
- For the small field size, the y-axis centring point was translated in the posterior direction, while the x- and z-axes centring points were left as they were.

Figure 3: Irradiated field (large size). The black dashed box represents CD16, while the red box represents older versions. The left image shows a PA view (i.e. looking from behind), while the right image shows a left lateral view. The lungs have been removed for clarity.

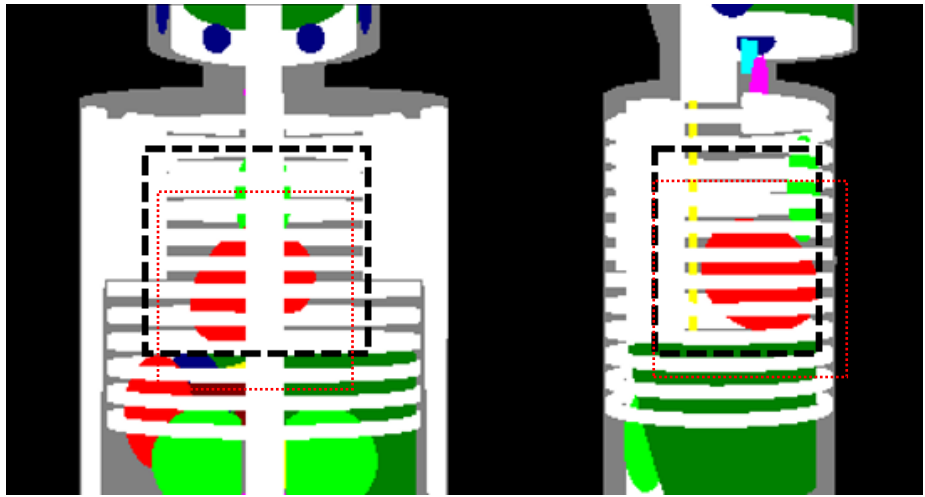

Figure 4: Irradiated field (small size).

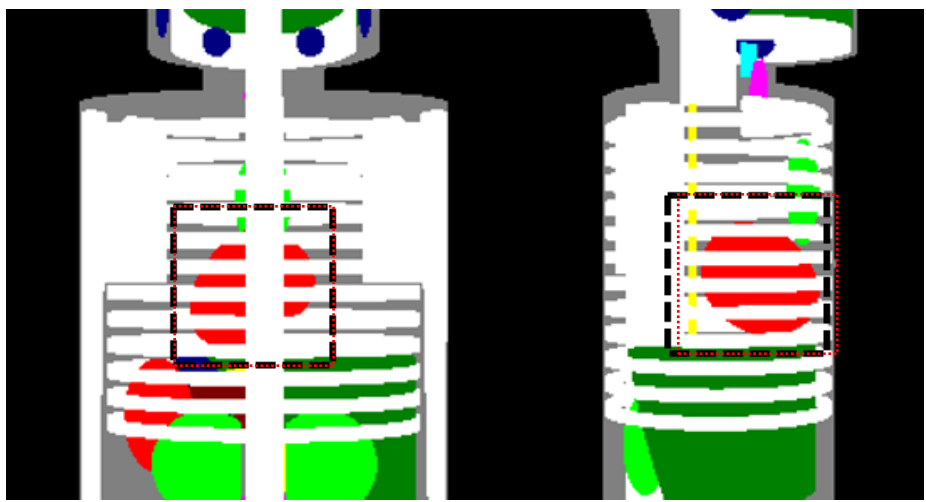

The second major change since earlier versions was the use of field size specific energy corrections. Previously, the energy correction factors calculated for the small field size were also used for the large field size, resulting in small errors. For the current version of CD16, each field size has its own set of energy correction polynomials.

Comparison with previous publications:

Tables 2 and 3 compare the effective dose per unit $P_{KA}$ (mSv/Gy*cm$^2$) obtained using CD16 with equivalent figures reported in other studies. These figures are reasonably consistent, although almost all dose estimates for paediatric cardiac catheterizations involve PCXMC, meaning variation is mostly due to different beam spectra, field size and centring point.

Table 2: Effective dose per unit $P_{KA}$ (mSv/Gy*cm$^2$) for the PA projection

| Study | Methodology | Beam energy | Phantom size | | | | | |
|---|---|---|---|---|---|---|---|---|
| | | | 0 y | 1 y | 5 y | 10 y | 15 y | Adult |
| CD16 Field size 1 | PCXMC (V2.0) | HVL=4.0 mm Al | 3.11 | 1.37 | 0.72 | 0.41 | 0.22 | 0.17 |
| CD16 Field size 1 | | HVL=5.0 mm Al | 3.84 | 1.62 | 0.86 | 0.49 | 0.28 | 0.21 |
| CD16 Field size 1 | | HVL=6.0 mm Al | 4.34 | 1.83 | 0.98 | 0.56 | 0.33 | 0.25 |
| CD16 Field size 2 | | HVL=4.0 mm Al | 3.13 | 1.40 | 0.78 | 0.50 | 0.26 | 0.21 |
| CD16 Field size 2 | | HVL=5.0 mm Al | 3.57 | 1.66 | 0.92 | 0.60 | 0.32 | 0.26 |
| CD16 Field size 2 | | HVL=6.0 mm Al | 3.92 | 1.87 | 1.05 | 0.69 | 0.37 | 0.30 |
| Barnaoui [7] | PCXMC (V2.0) | 70 kV, 3 mm Al, 0.3 mm Cu | 3.5 | 1.6 | 0.8 | 0.5 | 0.3 | - |
| Karembatsakidou | PCXMC (V2.0) | 60-85 kV, 6 mm Al, 0.2/0.4 mm Cu | 3.7 | 1.9 | 1 | 0.6 | 0.4 | - |
| Schmidt [8] [a] | PCXMC (V1.3) | 65 kV, 3.0 mm Al | 2.05 | 0.82 | 0.42 | 0.24 | 0.13 | 0.10 |
| Axelsson [9] | Physical measurements | 58-70 kV, 3.0 mm Al | - | 1.8 | 0.9 | - | - | - |
| Glatz [10] [b] | PCXMC | Not stated | 2.07 | 0.91 | 0.68 | 0.47 | 0.21 | 0.18 |
| Dragusin [11] | PCXMC | HVL=6.0 mm | 3.61 | 2.19 | 0.91 | 0.71 | 0.41 | |
| Steulens (5.5 mm Al HVL) [12] | MCNP | HVL=5.5 mm Al | - | - | - | - | - | 0.28 |
| Kawasaki [13] | Physical measurements | HVL=7.6/6.4 mm Al (0 y), 7.7, 6.7 mm Al (1 y) | 2.34/2.2 | 1.27/1.4 | - | - | - | - |

[a] ICRP 60 weighting factors

[b] Data are for patient mass ranges (<5, ≥5 to ≤12.5, ≥12.5 to ≤25, ≥25 to ≤45, ≥45 to ≤65 and ≥65 kg), rather than age.

Table 3: Effective dose per unit $P_{KA}$ (mSv/Gy*cm$^2$) for the left lateral projection:

| Study | Methodology | Beam energy | Phantom size | | | | | |
|---|---|---|---|---|---|---|---|---|
| | | | 0 y | 1 y | 5 y | 10 y | 15 y | Adult |
| CD16 Field size 1 | PCXMC (V2.0) | HVL=4.0 mm Al | 3.40 | 1.55 | 0.91 | 0.56 | 0.32 | 0.24 |
| CD16 Field size 1 | | HVL=5.0 mm Al | 4.05 | 1.79 | 1.06 | 0.66 | 0.38 | 0.29 |
| CD16 Field size 1 | | HVL=6.0 mm Al | 4.49 | 1.98 | 1.19 | 0.75 | 0.44 | 0.34 |
| CD16 Field size 2 | | HVL=4.0 mm Al | 3.11 | 1.49 | 0.84 | 0.51 | 0.30 | 0.21 |
| CD16 Field size 2 | | HVL=5.0 mm Al | 3.52 | 1.72 | 0.99 | 0.61 | 0.36 | 0.26 |
| CD16 Field size 2 | | HVL=6.0 mm Al | 3.82 | 1.90 | 1.11 | 0.70 | 0.41 | 0.30 |
| Barnaoui [7] | PCXMC (V2.0) | 70 kV, 3 mm Al, 0.3 mm Cu | 3.50 | 2.60 | 1.30 | 0.80 | 0.40 | - |
| Karembatsakidou | PCXMC (V2.0) | 60-85 kV, 6 mm Al, 0.2/0.4 mm Cu | 3.70 | 1.90 | 1 | 0.70 | 0.4 | - |
| Schmidt [8] [a] | PCXMC (V1.3) | 65 kV, 3.0 mm Al | 2.34 | 1.16 | 0.64 | 0.38 | 0.22 | 0.16 |
| Axelsson [9] | Physical measurements | 58-70 kV, 3.0 mm Al | - | 1.4 | 0.7 | - | - | - |
| Glatz [10] [b] | PCXMC | Not stated | 2.25 | 1.04 | 0.89 | 0.6 | 0.24 | 0.16 |
| Dragusin [11] | PCXMC | HVL=5.7 mm Al | 3.31 | 2.17 | 0.87 | 0.65 | 0.39 | |
| Steulens (5.5 mm Al HVL) [12] | MCNPX 2.5.0 | HVL=5.5 mm Al | - | - | - | - | - | 0.25 |
| Kawasaki [13] | Physical measurements | HVL=6.6 mm Al | 4.0 | 2.7 | - | - | - | - |

[a] ICRP 60 weighting factors

[b] Data are for patient mass ranges (<5, ≥5 to ≤12.5, ≥12.5 to ≤25, ≥25 to ≤45, ≥45 to ≤65 and ≥65 kg), rather than age.

### Limitations of CD16:

The Monte Carlo simulations from which the k-factors used in CD16 were derived involve crude Cristy/Eckerman phantoms with simplistic organ shapes (e.g. see Figures 3 and 4). The lungs are simple ellipsoid shapes, diverging away from the midsagittal plane toward the apex to a much greater extent than occurs in reality (Figure 5). The diaphragm is also flat, rather than dome-shaped in humans. Consequently, the superior portion of the liver and the fundus of the stomach may extend superiorly (and thus enter the primary x-ray field) more in reality, than suggested by the Christy phantom. Lung density is fixed at 0.30 g cm$^{-3}$ for all phantom sizes. In reality, growth of new alveoli ends at age 2 years. Beyond this age, lung growth involves increased size of existing alveoli, with a corresponding decrease in lung density with age [14].

Figure 5: Chest and upper abdomen region in PCXMC phantom

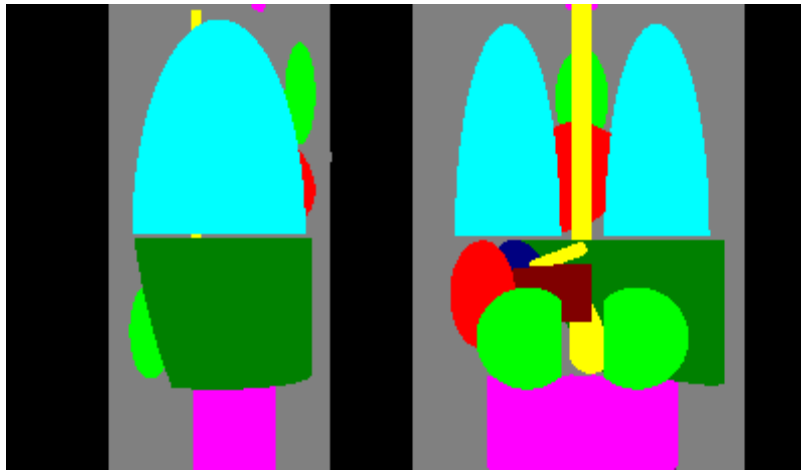

A further major shortcoming of the Christy/Eckerman phantoms is the unrealistic simulation of variation in mass for a given height. A simplistic algorithm is used resulting in all organs expanding like a balloon, bones and soft tissues alike (Figure 6). This has no basis in reality. Consequently, adjustments for obesity were omitted from CD16.

Figure 6: The bizarre effect of adjusting phantom mass for a given height in PCXMC

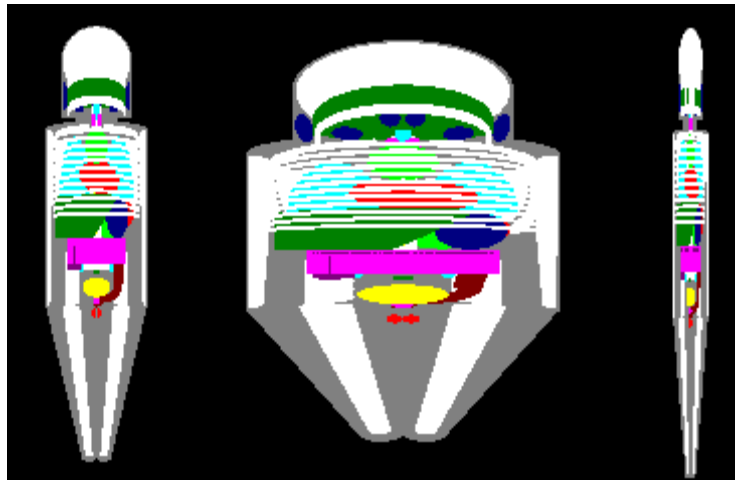

## Conclusions:

CD16 allows rapid estimation of organ doses from x-ray guided cardiac catheterization procedures. It is suitable for epidemiological studies in which information on historical exposures is often limited and sample sizes may be large. Future work will involve (1) calculating new *k*-factors using more realistic voxel phantom models, (2) a 2DMC approach to uncertainty modelling and (3) obtaining more data on beam angles and beam energy used in clinical practice.

## Appendix: List of versions

Cardiodose: original version using polynomials to encode dose variation with beam angles.

Cardiodose2: beam angle polynomials replaced by linear interpolation

Cardiodose3: added stomach dose

Cardiodose4: added whole body average dose

Cardiodose5: energy correction polynomials combined into a single file.

Cardiodose6: new fields, excluding breasts from primary beam in lateral projections. Single energy correction file for both field sizes

CD16: as Cardiodose6, but with field size-specific energy corrections.